\documentstyle[graphics,twocolumn]{mn}
\newcommand{\mincir}{\raise
-2.truept\hbox{\rlap{\hbox{$\sim$}}\raise5.truept 
\hbox{$<$}\ }}
\newcommand{\magcir}{\raise
-2.truept\hbox{\rlap{\hbox{$\sim$}}\raise5.truept
\hbox{$>$}\ }}
\newcommand{\minmag}{\raise-2.truept\hbox{\rlap{\hbox{$<$}}\raise
6.truept\hbox
{$>$}\ }}

\renewcommand{\H}{{\mbox{${\rm H{\sc i}~}$}}}

\newcommand{\Hep}{{\mbox{${\rm He{\sc ii}~}$}}}

\newcommand{\lya}{Lyman-$\alpha$~}
\newcommand{\gad} {{\small {GADGET-2}}\,}

\newcommand{\be}{\begin{equation}}
\newcommand{\ee}{\end{equation}}
\newcommand{\ba}{\begin{eqnarray}}
\newcommand{\ea}{\end{eqnarray}}
\newcommand{\brr}{\begin{array}}
 
\newcommand{\err}{\end{array}}
\newcommand{\bc}{\begin{center}}
\newcommand{\ec}{\end{center}}

\newcommand{\vel}{\,{\rm km\,s^{-1}}}

\DeclareMathAlphabet{\mathsc}{OT1}{cmr}{m}{sc}
\def\testbx{bx}%
\DeclareRobustCommand{\ion}[2]{%
\relax\ifmmode
\ifx\testbx\f@series
{\mathbf{#1\,\mathsc{#2}}}\else
{\mathrm{#1\,\mathsc{#2}}}\fi
\else\textup{#1\,{\mdseries\textsc{#2}}}%
\fi}
\title[Cosmological and astrophysical parameters from the SDSS flux power spectrum]
{Cosmological and astrophysical parameters from the SDSS flux power spectrum and
hydrodynamical simulations of the \lya forest}

\author[M. Viel \& M.G. Haehnelt] {Matteo Viel \&
Martin G. Haehnelt \\ Institute of
Astronomy, Madingley Road, Cambridge CB3 0HA\\
\\}

\begin{document}
\maketitle
\begin{abstract}
The flux power spectrum of the \lya forest in quasar (QSO) absorption
spectra is sensitive to a wide range of cosmological and astrophysical
parameters and instrumental effects. Modelling the flux power spectrum
in this large parameter space to an accuracy comparable to the
statistical uncertainty of large samples of QSO spectra is very
challenging.  We use here a coarse grid of hydrodynamical simulations
run with \gad to obtain a ``best guess'' model around which we
calculate a finer grid of flux power spectra using a Taylor expansion
of the flux power spectrum to first order. In this way, we investigate
how the interplay between astrophysical and cosmological parameters
affects their measurements using the recently published flux power
spectrum obtained from 3035 SDSS QSOs (McDonald et al. 2004).  We find
that the SDSS flux power spectrum alone is able to constrain a wide
range of parameters including the amplitude of the matter power
spectrum $\sigma_8$, the matter density $\Omega_{\rm m}$, the spectral
index of primordial density fluctuations $n$, the effective optical
depth $\tau_{\rm eff}$ and its evolution. The thermal history of the
Intergalactic Medium (IGM) is, however, poorly constrained and the SDSS
data favour either an unplausibly large temperature or an unplausibly
steep temperature-density relation.  By enforcing a thermal history of
the IGM consistent with that inferred from high-resolution QSO spectra,
we find the following values for the best fitting model (assuming a
flat Universe with a cosmological constant and zero neutrino mass):
$\Omega_{\rm m}=0.28 \pm 0.03$, $n=0.95\pm0.04$, $\sigma_8=0.91\pm0.07$
($1\sigma$ error bars). The values for $\sigma_8$ and $n$ are consistent with those obtained
by McDonald et al. with different simulations for similar
assumptions. We argue, however, that the major uncertainties in this
measurement are still systematic rather than statistical.
\end{abstract}


\section{Introduction}

The last few years haven seen the establishment of the \lya forest as
one of the major observational tools to probe the matter power
spectrum.  Measurements of the matter power from the \lya forest data extend to 
smaller scales and probe a complementary redshift range than those
using the Cosmic Microwave Background (CMB),
galaxy surveys or weak gravitational lensing. The \lya forest
 is thus ideally suited to constrain cosmological
parameters which affect the power spectrum 
on small scales like the neutrino mass and the  mass of Dark Matter
(DM) particles (Seljak et al. 2004; Viel et al. 2005). 
In a combined analysis with CMB data the \lya forest results have also
been pivotal in tightening constraints on the power law index of the 
power spectrum of primordial density fluctuations $n$.  

A consistent picture emerges suggesting that the fluctuation amplitude
of the matter power spectrum $\sigma_8$ is rather high, that the
spectral index of primordial density fluctuations is consistent with the
value $n=1$ and that there is no evidence for a running of
the spectral index, significant neutrino mass or a deviation from a
cold dark matter spectrum at small scales (Croft et al. 1998, McDonald
et al. 2000, Hui et al. 2001, Croft et al. 2002 [C02], McDonald 2003,
Viel et al. 2003, Viel, Haehnelt \& Springel 2004 [VHS], Desjacques \&
Haehnelt 2004, Viel, Weller \& Haehnelt 2004, McDonald et al. 2004
[M04a], McDonald et al. 2004b [M04b], Viel et al. 2005, Lidz et
al. 2005).  Major uncertainties are thereby the assumed effective
optical depth and thermal history of the Intergalactic Medium (IGM) and
the numerical limitations in obtaining accurate theoretical predictions
for the flux power spectrum for a large parameter space (see VHS and
M04a for a
detailed discussion).

VHS and Viel, Weller \& Haehnelt (2004) recovered the linear dark
matter power spectrum amplitude and its slope from a new set of 27 high
resolution high signal-to-noise QSOs (the LUQAS sample, Kim et al. 2004
[K04]) and reanalysed the earlier results of C02.  Viel et al.  found
$\sigma_8=0.94\pm0.08$, $n=0.99\pm0.03$ ($1\sigma$) and no evidence for
a (large) running spectral index when they combined the \lya forest
data with the WMAP data.  Similar results have have been 
obtained subsequently  by the SDSS
collaboration from a much larger sample of 3035 low-resolution low S/N spectra
with significantly wider reshift coverage 
(M04b; Seljak et al. 2004). These findings corrected earlier claims of
Spergel et al. (2003) who used the data of C02 combined with CMB and
galaxy survey data to argue for a significant tilt away from the
canonical Harrison-Zeldovich $n=1$ spectrum and possibly also for a
running of the spectral index (see Seljak et al. (2003) for the first
suggestion that this result may be due to the assumption of an
unplausibly large effective optical depth).

Large samples of QSO absorption spectra offer the opportunity to obtain
the flux power spectrum for a wide range of redshifts with an accuracy
at the percent level. Despite the generally good agreement between
different groups achieved in the last couple of years there are,
however, still major open issues of what and how large the
uncertainties in the cosmological and astrophysical parameters inferred
from the flux power spectrum are (see VHS and M04b for discussions with
different views). It will be important to resolve these issues if
further progress is to be made.

The two data sets used by VHS and M04b and their theoretical modelling
were very different.  The SDSS QSO data set analysed by M04a consists
of $3035$ QSO spectra at low resolution ($R\sim 2000$) and low S/N
($\sim 10$ per pixel) spanning a wide range of redshifts, while the LUQAS
and the C02 samples contain mainly high resolution ($R \sim 45000$),
high signal-to-noise ($>50$ per pixel) QSO spectra with median
redshifts of $z=2.125$ and $z=2.725$, respectively. The analysis
methods to infer cosmological and astrophysical parameters were also
very different.  M04b modelled the flux power spectrum using a large
number of HPM simulations (Gnedin \& Hui 1998; Viel, Haehnelt \&
Springel 2005) exploring a large multi-dimensional parameter
space. Viel et al. (2004) improved instead the effective bias method
developed by C02.
They used a grid of full hydrodynamical simulations
run with the Tree-SPH code \gad (Springel, Yoshida \& White 2001; Springel 2005) to
invert the non-linear relation between flux and matter power spectrum.

Both methods have their own set of problems.  M04b had to use the
approximate HPM method which they could only calibrate on a small
number of hydro simulations of small box size (see Viel, Haehnelt \&
Springel (2005) for a discussion of the accuracy of HPM simulations).
M04b further had to compromise on box size and even time-stepping in
order to explore a large parameter space. This lead to large number of
uncertain corrections which added to another large set of corrections
necessary because of the rather low resolution and S/N of their
data. The method used by VHS on the other hand requires differentiating
the 1D flux power spectrum to obtain the ``3D'' flux power spectrum,
which introduces rather large errors. This makes the assessment of
statistical errors somewhat difficult. 
We refer to Zaldarriaga, Hui \& Tegmark (2001); Gnedin \& Hamilton (2002);
Zaldarriaga, Scoccimarro \& Hui (2003) and Seljak, Mcdonald \&Makarov
(2003) for a further critical assessments of the  use of the effective bias
method to infer cosmological parameters. VHS concluded that the best they
can do was to give an estimate of the many systematic uncertainties
involved and to combine these in a conservative way in the final
result. However, for attemtps to take advantage of  the full redshift 
coverage and smaller errors of the observed 
SDDS flux power spectrum the effective bias stops being useful.

Further motivated by the differences in the data and in the
theoretical modelling performed
by the two groups, we will analyse here the SDSS flux power spectrum
with  high resolution, large box-size full hydrodynamical
simulations. Ideally, one would like to be able to repeat the analysis
made by the SDSS collaboration by using a very fine grid of full
hydrodynamical simulations in order to be able to sample the 
multi-dimensional parameter space. However, at present, this approach is not feasible 
since hydrodynamical simulations are very time consuming.
We therefore decided to concentrate most of the analysis on a small
region of the parameter space, around the best-fit models obtained by
the two groups. We thereby use a Taylor
expansion to approximate the flux power spectrum around a best-guess 
model. This approximation should be reasonably accurate for little
displacements in parameter space. We have checked the  validity
of the approximation for a few models. However, we caution the reader 
that the final error  bars on recovered astrophysical and cosmological 
parameters is likely  to be somewhat underestimated due to unaccounted 
errors of this approximation.

The plan of this paper is as follows. In Section \ref{data} we briefly
describe the SDSS data set, while the hydrodynamical simulations
are discussed in Section \ref{hydro}.  Section \ref{results} describes
the technical details of our modelling of the flux power spectrum 
and the reader more interested in the results may go straight to
Section \ref{chifitting}  where we discuss our findings for the
cosmological and astrophysical parameters. 
Section \ref{conclu} contains a summary and an outline of possible ways
of improving these measurements.

\section{The  SDSS flux power spectrum}
\label{data}

McDonald et al. (2004a) have presented the flux power spectrum 
of a large sample of 3035 absorption spectra of QSO in the redshift 
range $2<z<4$ drawn from the DR1 and DR2 data releases of SDSS.  
With a spectral  resolution of   $R\sim 2000$ the typical absorption 
features with a width of $\sim 30$ km/s are not resolved.  The
signal-to-noise of the individual spectra  is rather low, S/N  $\sim 10$ 
per pixel. The large number of spectra means, however, that the flux
power spectrum on scales a factor of a few larger than the 
thermal cut-off can be measured with small statistical errors. 
M04a have re-analyzed the raw absorption spectra 
and have investigated the effect of noise, resolution of the
spectrograph, sky subtraction, quasar continuum and  associated
metal absorption. M04a make corrections for these effects and give
estimates of the errors for the most important of these corrections. 
The corrections are not small. The noise contribution to the flux
power spectrum rises from 15-30 percent at the smallest wavenumbers 
to order unity at the largest wavenumbers and varies with redshift. 
The correction for uncorrelated metal absorption are generally a
factor five  to ten smaller than this. The correction for resolution 
varies from 1\% at the smallest wavenumbers to a factor four at the 
largest wavenumbers. M04a also identified a correlated 
SiIII feature in the absorption spectra for which they present an 
empirical fit to the effect on the flux power spectrum. As final
result of their analysis M04a present their best estimate for the 
flux power spectrum $P_F(k,z)$ at 12 wavenumbers in the range
$0.00141<k\,$(s/km)$<0.01778$, equally spaced in $\Delta \log k = 0.1$
for $z=2.2,2.4,2.6,2.8,3,3.2,3.4,3.6,3.8,4,4.2$. At the time we started
this work the two highest redshift bins were not publicly available so we
will not use them in the following analysis. 
We will here use this flux power spectrum together with the recommended
corrections to the data and the recommended treatment of the 
errors of  these corrections. We will come back to this in 
more detail in section \ref{datacorr}. 
Note that we have  also dropped the highest redshift bin at
$z=3.8$ as we could not fit it well with our models for the flux 
power spectrum  (see section \ref{bestguess} for more details).

\section{hydrodynamical simulations}
\label{hydro}

\begin{table}
\caption{Grid of hydrodynamical cosmological simulations}
\label{tab1}
\begin{tabular}{lccccc}
\hline
\noalign{\smallskip}
${\rm simulation}$ &  ${\rm\sigma_8}$ & ${\rm n}$ &
${\rm box}-{\rm {N_p}^{1/3}}$ & ${\rm \chi^2_{min}}^{\mathrm{(a)}}$\\
\noalign{\smallskip}
\hline
\noalign{\smallskip}
B1    & 0.7    & 0.95  & 60-400& 124.1\\
B1.5  & 0.775  & 0.95  & 60-400 &113.2\\
B2    & 0.85   & 0.95  & 60-400 & 104.7\\
B2.5  & 0.925  & 0.95  & 60-400 & 101.2\\
B3    & 1      & 0.95  & 60-400 & 107.7 \\
B3.5  & 1.075  & 0.95   & 60-400&132.8 \\
C2    & 0.85   & 1     & 60-400 &127.8\\
C3    & 1      & 1     & 60-400 &121.2\\
B2$_{\rm 30,200}$  & 0.85    & 0.95  & 30-200 & --\\
B2$_{\rm HO}^{\mathrm{(b)}}$ & 0.85    & 0.95  & 30-200& -- \\
B2$_{\rm \Omega_m}^{\mathrm{(c)}}$ & 0.85    & 0.95  & 30-200& --\\ 
B2$_{\rm ne}\,^{\mathrm{(d)}}$ & 0.85    & 0.95  & 30-200& --\\ 
B2$_{\rm nec}\,^{\mathrm{(e)}}$ & 0.85    & 0.95  & 30-200& --\\
B2$_{\rm lr}\,^{\mathrm{(f)}}$ & 0.85    & 0.95  & 30-200& --\\ 
B2$_{\rm hr}\,^{\mathrm{(g)}}$ & 0.85    & 0.95  & 30-200& --\\ 
B2$_{\rm cold}\,^{\mathrm{(h)}}$ & 0.85    & 0.95  & 30-200& --\\ 
B2$_{\rm 30,400}$ & 0.85    & 0.95  & 30-400 & -- \\ 
B2$_{\rm 120,400}$ & 0.85    & 0.95  & 120-400& --\\ 
\hline
\noalign{\smallskip}
\end{tabular}
\begin{list}{}{}
\small
\item[$^{\mathrm{(a)}}$] $\chi^2_{\rm min}$ is for 88 d.o.f.
(96 data points, eight free parameters for the effective 
optical depth at the eight different redshifts) errors for the noise and resolution 
correction are treated as suggested in M04a;
$^{\mathrm{(b)}}$ $H_0=80$ km/s/Mpc; $^{\mathrm{(c)}}$ 
$\Omega_{\rm m}=0.30$;
$^{\mathrm{(d)}}$ simulation with non-equilibrium solver with reionization at $z \sim
6$; $^{\mathrm{(e)}}$ non eq. colder version with reionization at $z
\sim 6$; $^{\mathrm{(f)}}$ non eq. version with late reionization at $z
\sim 4$; $^{\mathrm{(g)}}$ non eq. version with early reionization at
$z \sim 17$; $^{\mathrm{(h)}}$ same as $B2_{\rm 30,200}$ (equilibrium with
reionization at $z \sim 6$) but with colder equation of state to match
the evolution of B2$_{\rm hr}$ at $z<4$.
\end{list}
\end{table}

We have run a suite of  full hydrodynamical simulations with \gad 
(Springel, Yoshida \& White 2001; Springel 2005)
similar to those in VHS. We have varied the  cosmological
parameters, particle number, resolution, box-size and thermal history
of the simulations. In Table 1 we list the fluctuation amplitude 
$\sigma_{8}$, the spectral index $n$ as well as box size and number 
of particles of the different simulations (note that few  of these
simulations are actually the same as those presented in VHS). The box-size and particle
number are given in a form such that 60-400 corresponds to a box 
of length $60 h^{-1}\,$ Mpc with  $2\times 400^{3}$  (gas + DM) particles. 
Note that this is the box size of the  simulations used for our final 
analysis and is a factor 6 larger than the largest of the hydrodynamical
simulations used by M04b. 

\gad was used in its TreePM mode and we have used a simplified 
star formation criterion which speeds up the calculations
considerably. The simulations were performed with periodic boundary 
conditions with an equal number of dark matter and gas particles and 
used the conservative `entropy-formulation' of SPH proposed by 
Springel \& Hernquist (2002). The cosmological parameters are close to the
values obtained by the WMAP team in their combined analysis of CMB and other
data (Spergel et al. 2003).  All but two of the simulations in  Table 1 
have the following parameters: $\Omega_{{\rm m}}= 0.26$, $\Omega_{\Lambda}
= 0.74$, $\Omega_{{\rm b}} = 0.0463$ and $H_0=72\,{\rm
km\,s^{-1}Mpc^{-1}}$.  The CDM transfer functions of all models have
been taken from Eisenstein \& Hu (1999). For the remaining two simulations the 
Hubble constant and matter density were varied to  
 $\Omega_{{\rm m}}=0.3$ and $H_0=80\,{\rm km\,s^{-1}Mpc^{-1}}$,
respectively.  
To facilitate a comparison with M04b we also give the amplitude of the linear power
spectrum at the pivot wavenumber $k_p$ used by M04b, $\Delta^2_L(k_p=0.009\,{\rm s/km},z_p=3)=0.350$ and 
the effective spectral index $n_{\rm eff}=d \ln P_L/ d\ln k=-2.33$ for
the B2 simulation.

Most of the simulations were run with the {\it equilibrium} solver 
for the evolution of the ionization balance and temperature
implemented in the public version of \gad  (which  assumes the gas to 
be in photo-ionization equilibrium) with a UV background produced by
quasars as given by Haardt \& Madau (1996), which leads to
reionisation of the Universe at $z\simeq 6$. 
The assumption of photoionization equilibrium is valid for most of the 
evolution of the IGM responsible for the \lya forest. It is, however, 
a bad approximation during reionization, where it leads to an
underestimate of the photo-heating rate and as a
result to too low temperatures and generally too steep a temperature density
relation (Theuns et al. 1998).  For the models run with the equilibrium
solver we have thus as before increased the heating rates by a factor of 3.3 at $z>3.2$
in order to take into account the underestimate of the photo-heating due
to the equilibrium solver and optical depth effects for the photo-heating of
helium before helium is fully reionized at $\sim 3.2$ (Abel \& Haehnelt 1999). 
This is necessary to obtain temperatures  close to  observed
temperatures (Schaye et al. 2000, Ricotti et al. 2000). 

Some simulations were run with a {\it non-equilibrium} solver 
which has been implemented by James Bolton into \gad.  
For the simulations run with the non-equilibrium solver 
only the optical depth effect for the photoheating of helium have 
to be taken into account. The  increase of 
the photo-heating rates at $z>2$  necessary to match observed temperatures was 
thus only a factor 1.8. We have varied the thermal history of the 
simulations by re-mapping the redshift evolution of the UV
background. For more details on the simulations and the
non-equilibrium solver we refer to VHS, Bolton et al. (2005a, 2005b)
and section \ref{thermal}.

\section{Modelling the SDSS flux power spectrum using hydro-simulations} 
\label{results}

\subsection{A two-step approach}

Hydro simulations are rather expensive in terms of CPU time. A typical
simulation used in the analysis of VHS took 2 weeks of wall-clock time
to reach $z=2$ on 32 processors on COSMOS, an SGI Altix 3700. This
made it  impossible to fully sample a large multi-dimensional parameter 
space with full hydro-simulations. With the rather modest numerical resources
available to us we have thus  decided to take the following approach. 

We have first run a range of simulations with
cosmological parameters close to those inferred by  VHS and
M04b which allowed us to explore a wider range of thermal histories
of the IGM to better understand the resulting uncertainties. 
We have then fitted the simulated flux power spectra to the 
SDSS flux power spectrum as described in section \ref{chifitting}. 
Based on this and our previous studies in VHS we have chosen 
a ``best guess'' model which fits the data well. To improve our
parameter choices further and  to get at least approximate 
error estimates we have then explored a large multi-dimensional parameter
space around our best-guess model.  This also enabled us to 
explore  the degeneracies between different parameters.  
In order to keep the required CPU time at a manageable level we have
approximated the flux power spectra by a Taylor expansion to first
order around our best guess model. If ${\bf p}$ is an arbitrary
parameter vector close to the best guess model described by ${\bf
p_0}$, we have assumed that:

\begin{eqnarray}
P_F(k,z;{\bf p})& & =  P_F(k,z;{\bf p^0})  + \nonumber \\
& &
+\sum_i^N \frac{\partial {P_F(k,z;p_i)}}{\partial {p_i}}\bigg|_{\bf
{p}=\bf{p^0}} (p_i - p_i^0)\, ,
\label{taylor}
\end{eqnarray}
where $p_i$ are the N components of the vector ${\bf p}$. 
This requires
the determination of only N derivatives which can be estimated by
running one simulation each close to the best guess model.  Obviously
this linear approximation will only hold as long as the changes in the
flux power spectrum are small. Within 1 to 2 $\sigma$ of the  best
guess model this should, however, be the case. We have explicitly
checked how these derivatives change around another fiducial model for
some of the parameters and we will quantify this in section \ref{derivatives}.
Once the derivatives are obtained we can then calculate an arbitrarily
fine grid of flux power spectra around the best guess model
with equation \ref{taylor}.

\subsection{Systematic Uncertainties}
\label{system}

As discussed in detail by VHS and M04b there is a wide range of
systematic uncertainties in the analysis of the flux power spectrum. 
The origins of these uncertainties fall broadly into five categories: 
\begin{itemize} 
\item{deficiencies of the data which have to be corrected;} 
\item{uncertainty of the effective optical depth;} 
\item{uncertainty of the thermal state of the IGM;} 
\item{lack of ability to make accurate 
      predictions of the flux power spectrum for a large parameter 
      space;} 
\item{lack of ability to model other physical
      processes which potentially affect the flux power spectrum.}
\end{itemize} 
We will discuss  the resulting systematic uncertainties in  turn. 

\subsubsection{Correction to the data}
\label{datacorr}

\begin{figure}
\center\resizebox{0.5\textwidth}{!}{\includegraphics{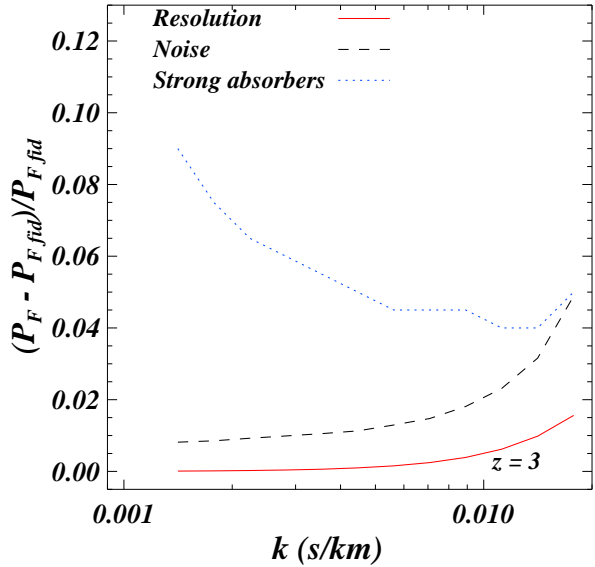}}
\caption
{The dashed, solid and dotted curves show the effect of the error in 
the noise correction and resolution  and the correction for strong
absorption systems, respectively. We show fractional differences in
the flux power spectrum at $z=3$.  Note that we assume only the error 
of the noise correction  to depend on redshift. The changes correspond
to an error  of 5\% in the noise correction and of (7 km/s)$^2$ in
the resolution (see section \ref{datacorr} for more details).}
\label{f1}
\end{figure}

M04b have applied a number of corrections for noise, resolution and
the correlated SiIII feature identified by M04a. We have adopted 
the same corrections which are summarized below. For more details 
we refer to M04a and M04b. We allow for an error in the 
$k-$dependent noise correction at each redshift by subtracting 
$f_i\,P_{\rm noise}\,(k,z_i)$ from $P_F(k,z)$ and treat the $f_i$ as 
free parameter in the fit. M04b recommend to assume that the
distribution of $f_i$ is  Gaussian with mean zero and width 0.05 which corresponds to 
a typical error in the noise correction of $5\%$.  M04a further
recommend to allow for an overall error on the
resolution correction, by multiplying $P_F(k,z)$ with
$\exp(\alpha\,k^2)$, where 
$\alpha$ is treated as a free parameter with Gaussian
distribution with mean zero and width $(7\vel)^2$.  To account for contamination by  
the correlated SiIII feature we modify our model  flux power spectra 
as suggested as follows, $P_F(k)=(1+a^2)\,P_{\rm sim}(k)+ 2\,a\cos(v\,k)\,P_{\rm sim}(k)$
with  $a=f_{\rm SiIII}/[1-\overline{F}(z)]$, $f_{\rm SiIII}=0.11$ and $v=2271$
km/s. 
The noise and resolution corrections and their
errors depend strongly on the wavenumber $k$ as shown in  
Figure \ref{f1}. Note that M04a also
subtracted a contribution of uncorrelated associated metal 
absorption which they estimated redward of the \lya emission 
line.  The effect  of continuum fitting errors on the flux power
spectrum is a further uncertainty which is discussed in K04, M04a and Tytler et al. (2004).  
Continuum fitting errors are difficult to quantify and we again 
follow M04b in not attempting to model them.

\subsubsection{Uncertainty of the effective optical depth}
As discussed in VHS and Seljak et al. (2003) the  poorly known effective
optical depth results in major uncertainties in any analysis of the
\lya forest flux power spectrum. The effective optical depth is 
to a large extent degenerate with the amplitude of the matter power
spectrum. As shown in M04b the $k-$dependence of 
changes in the flux power spectrum due to varying effective  optical depth 
and amplitude of the matter power spectrum are, however, somewhat
different. This allows -- at least in principle -- to break this
degeneracy. We will come back to this point later. 
When modelling the effective optical depth we will investigate 
two cases.  We will either let the effective optical depth vary independently 
in the different redshift bins or we will
parameterize the evolution of the effective optical depth as a 
power law in redshift. 

\subsubsection{Uncertainties due to the thermal state of the IGM}
\label{thermal} 
\begin{figure*}
\center\resizebox{1.02\textwidth}{!}{\includegraphics{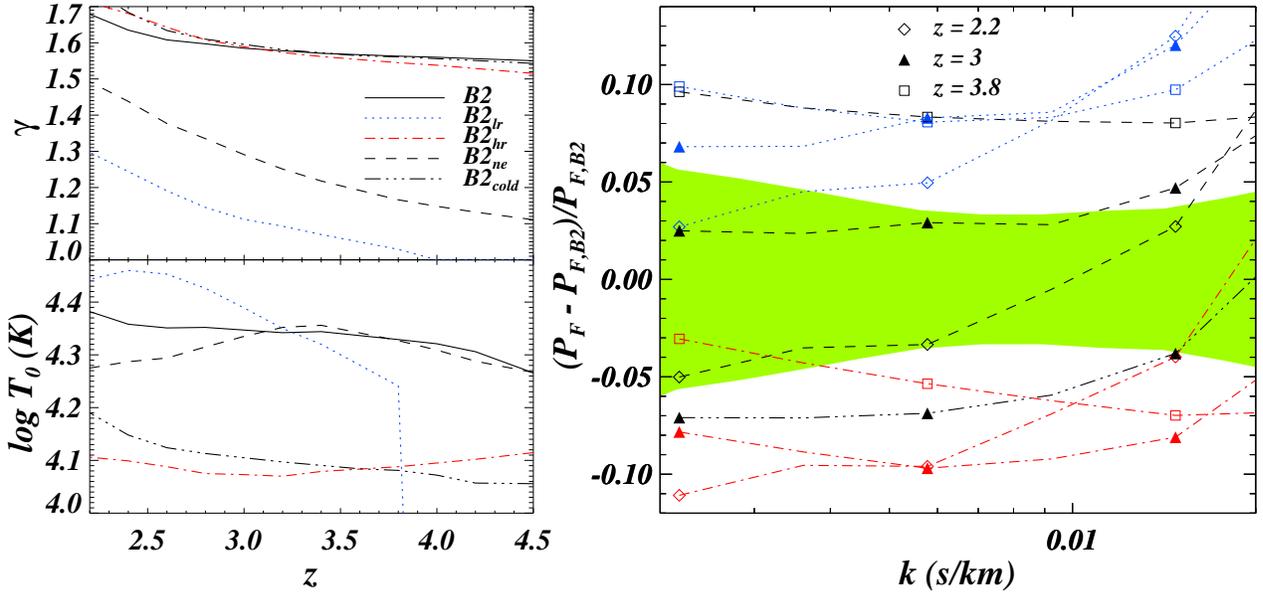}}
\caption
{{\it Left (bottom panel)}: 
Evolution of $T_0$ with redshift for different thermal
histories: B2, the fiducial equilibrium model with reionization at $z
\sim 6$ (continuous curve); B2$_{\rm lr}$, simulation with a non- equilibrium
solver with late reionization at $z\sim
4$ (dotted curve); B2$_{\rm hr}$, non-equilibrium, with early reionization at $z\sim
17$ (dot-dashed curve); B2$_{\rm ne}$, non-equilibrium
with reionization at $z\sim
6$ (dashed curve); B2$_{\rm cold}$, equilibrium
with reionization at $z\sim
6$ and with a colder equation of state (triple dotted-dashed curve).
{\it Left (top panel)}: Evolution of $\gamma$ with redshift. 
{\it Right}: Fractional differences between the flux power spectrum 
of the different model compared to the fiducial model  B2 at 
$z=2.2,3,3.8$ (diamonds, triangles and squares, respectively). For 
B2$_{\rm cold}$ differences are only shown for  $z=3$, the other
two redshifts are  very similar. The shaded region indicates the
statistical errors at $z=3$.}
\label{f3}
\end{figure*}

In order to explore in more detail the uncertainties due to thermal
effects on the flux power spectrum we have run simulations with a wide
range of thermal histories (Table 1). Note that this is different from
VHS where we have only performed a-posteriori rescalings of the
temperature-density relation to investigate thermal effects.  
We have calculated the value $T_0$ as the median
temperature of gas with logarithmic overdensity values between -0.1
and 0.1. The value of $\gamma$ we determined from the slope of the 
temperature-density relation for  gas at mean density and  1.1 times
the mean density.

In the
left panel of Figure \ref{f3} we plot the different thermal histories
(temperature at mean density in the bottom panel and $\gamma$ in the
top panel). The linestyles denote the different models as follows. The
continuous curve is our standard B2 model with the equilibrium
solver. The dashed curves shows the same model with the non-equilibrium
solver ($B2_{\rm ne}$) which results in a significantly smaller
$\gamma$.  The dotted and dot-dashed and curve shows the same model
with the non-equilibrium solver for late reionization at $z\sim 4$
($B2_{\rm lr}$) and early reionization at $z \sim 17$ ($B2_{\rm
hr}$). Finally, the dashed-triple-dot curves shows a model with
reionization at $z\sim 6$ but smaller heating rate and thus a lower
temperature, similar to that of $B2_{\rm hr}$.

In the right panel we show the differences of the flux power spectrum
relative to our fiducial model $B2$ at three different redshifts.  The
differences are typically $\sim 5-10$ \%.  Comparing $B2$ and $B2_{\rm
hr}$, which have a similar redshift evolution we find for example that a
decrease of $T_0$ by 50\% ($\sim 10000$ K) results in a decrease in the
flux power larger than $5 \%$. The effect of changing $\gamma$ can be
judged by comparing $B2_{\rm ne}$ and $B2_{\rm lr}$ at $z=3$ which have
approximately the same temperature. Decreasing $\gamma$ by $\Delta
\gamma = -0.5 (-0.3)$ produces an increase in the flux power by 7.5 \%
(2.5 \%), but we note that for $k>0.01$ s/km the differences increase
strongly.

We will later model the evolution of the thermal state of the IGM with
redshift as broken  power laws for temperature and slope of the temperature-density
relation.

\subsubsection{Uncertainties due to the modelling of the flux power spectrum}
\label{moderrors}

Modelling the flux power spectrum accurately is numerically challenging
and checks for convergence are important. We have chosen to use here
full hydro-simulations performed with \gad described in section
\ref{hydro}.  In VHS we found that simulations with a box size 60 $h^{-1}$
Mpc and with $2\times 400^3$ particles (60,400) are the best
compromise for the analysis of the LUQAS sample.  The lower resolution
SDSS power spectrum is not affected on large scale in the same way 
by continuum fitting errors  as  the high-resolution spectra which we
used in VHS and for which continuum fitting errors appear to increase
the flux power significantly at and above the scale corresponding to one 
Echelle order (Kim et al. 2004). 
The usable range of wavenumbers is thus shifted to larger scales
compared to  flux power spectra obtained from high-resolution
Echelle data.  Ideally one would thus like a somewhat larger box size
than that used in VHS. Unfortunately this is currently not feasible  with
simulations that at least marginally resolve the Jeans length.   
We have thus decided to use simulations with the same box size and 
resolution as in VHS.

{\it Box size }: For the final analysis we use simulations of 60$h^{-1}\,$
Mpc and with $2\times 400^3$ particles (60,400).  Since the error
bars for the observed 1D flux power spectrum 
are significantly smaller than those of the 3D flux power used by VHS 
we checked  whether or not the flux power spectrum is affected by box size
and/or limited resolution at the percent level.  
We have first  checked how well the largest scales probed by SDSS are
sampled by our fiducial (60,400) simulations. In VHS this was not an
issue  since the first data point used in the analysis was at
$k=0.003$ s/km, which was well sampled by the (60,400) simulations. 
Our analysis here  extends however to   $k=0.001$ s/km and could 
be affected by cosmic variance. To test this, we have compared the
flux power spectrum  of our fiducial (60,400) simulation B2 with that of 
a (120,400) simulation of the same model (B2$_{\rm 120,400}$). The power of 
the latter at  $k<0.003$ s/km is typically larger by $\sim 3,4,9$\% at
$z=2.2,3,3.8$, respectively. We correct for this by multiplying the
flux power spectrum of the (60,400) simulation by corresponding 
factors. We will later use 
simulations with a factor two smaller box size but the same resolution as
our fiducial simulation  to explore the effects of the thermal history. 
We have thus also compared the flux
power spectrum of our fiducial (60-400) simulation B2 with a (30-200) simulation  
of the same model (B2$_{\rm 30,200}$) and find that the latter
underestimates the power  by $\mincir 3\%$ in the range $0.003<k$ 
(s/km)$<0.03$ at $z=2.2,3,4$. Our findings regarding the dependence of the 
flux power spectrum on the thermal history of the IGM should thus be
little affected by the fact that they were obtained from simulations of
smaller box size.

{\it Resolution }: To quantify the effect of resolution, we compare
the simulation B2$_{\rm 30,200}$ with B2$_{\rm 30,400}$ a simulation
which has the same box sixe but eight times better mass resolution. The
differences between B2$_{\rm 30, 400}$ and B2$_{\rm 30,200}$ are less than 5 \% for
$k<0.01$ (s/km), at any redshift, (3\% in the range $0.004 < k $(s/km)
$<0.01$). At $k=0.02$ s/km, the flux power spectrum of  B2$_{\rm
30,400}$ is larger by $\sim 4\%, 12\%$, at $z=3$ and $z=4$,
respectively. This is in agreement with the findings of M04b. We will
correct for the limited resolution by multiplying the flux power
spectrum of our fiducial (60,400) simulation by corresponding factors.
Both the resolution and box size errors are comparable or smaller than the
statistical errors of the flux power spectrum. We explicitly checked that these
corrections do not significantly affect the parameters of our
best-fitting model. They do, however, affect the $\chi^2_{\rm min}$
value of the best-fitting model. Note that both the
resolution and box size corrections could be model dependent.

{\it High Column density systems/Damped \lya systems:} We have recently
pointed out that the absorption profiles of high column density
systems/damped \lya systems have a significant effect on the flux power
spectrum over a wide range of scales (Viel et al. 2004b).  These
absorption systems are caused by high-redshift galaxies and the gas in
their immediate vicinity. Numerical simulations still struggle to
reproduce these systems correctly and they are thus a major factor of
uncertainty in modelling the flux power spectrum. M04b have modelled
the expected effect on the SDSS flux power spectrum and we have
included this correction using the k-dependence of Fig. 11 in M04b.
M04b recommend to assume that the distribution of the correction made
$A_{\rm damp}$ is Gaussian with mean 1 and width 0.3.  The correction
is shown in Figure \ref{f1} as the dotted curve.  It is of order of
$10$\% at the largest scales and drops to 3\% at the smallest
scales. Note that unlike M04b we have assumed that this correction does
not vary with redshift.

\subsubsection{Uncertainties due to  UV fluctuations, galactic winds,
reionization history, temperature fluctuations}
\label{othereffects}

There are a range of physical processes some of which are difficult to
model, that are described below. Note that we have chosen not
to try to include these effects in our analysis.

{\it Spatial fluctuations of the \H ionization rate}: The flux power
spectrum is obviously sensitive to the neutral fraction of hydrogen
which depends on the \H ionization rate. At low redshift $z<3$ the mean
free path of ionizing photons is sufficiently large that spatial
fluctuations of the \H rate should be too small to affect the flux
power spectrum relevant for our investigation here (Meiksin \& White
2004; Croft 2004; McDonald et al. 2005).  At higher redshift this is
less obvious. Quantitative modelling of the fluctuations requires
however a detailed knowledge of the source distribution of ionizing
photons which is currently not available.  As we will discuss below
there may be some tentative evidence that UV fluctuations do become
important at the highest redshifts of the SDSS flux power spectrum
sample.

{\it Galactic Winds}: There is undeniable observational evidence that
the \lya forest has been affected by galactic winds. Associated metal
absorption is found to rather low optical depth (Cowie et al. 1995,
Schaye et al. 2003).  Searches for the effect of galactic winds on QSO
absorption spectra from Lyman break galaxies close to the line-of-sight
have also been successful. The volume filling factor of galactic winds
and the material enriched with metals is, however, very uncertain
(e.g. Pieri \& Haehnelt 2004; Adelberger et al. 2005; Rauch et
al. 2005).  Numerical simulations have generally shown the effect of
galactic winds to be small (Theuns et al. 2002a; Kollmeier et al. 2003;
Desjacques et al. 2004; McDonald et al. 2005; Kollmeier et al. 2005).

{\it Temperature fluctuations}:

It is  rather difficult to measure mean values of the 
temperature and little is known observationally about spatial 
fluctuations of the temperature (Theuns et al. 2002b). 
In the redshift range of interest the heating rate of the IGM 
should be dominated by photo-heating of helium before helium is fully
reionized.  Helium reionization should thus lead  to
spatial temperature fluctuations (Miralda-Escud\'e et al. 2000),
which will affect the neutral fraction of hydrogen through the
temperature dependence of the recombination
coefficient. Temperature fluctuation may thus affect the flux power spectrum. 
Quantitative modelling of their  effect on the flux power spectrum 
requires numerical simulation of  helium reionization including  full
radiative transfer and  good knowledge of the spatial
distribution of the source of \Hep ionizing photons. 
Such modelling will also be uncertain.  

{\it Reionization history}: The flux power spectrum is not only
sensitive to the current thermal state of the IGM but also to its past
thermal history. This is because the spatial distribution of the gas is
affected by pressure effects (Hui \& Gnedin 1997; Theuns, Schaye \&
Haehnelt 2000; Zaldarriaga, Scoccimarro, Hui 2001). Quantitative
modelling of this effect is, however, again very uncertain as the
heating rate at high redshift is expected to be dominated by
photo-heating of helium. In order to get a feeling for the effect of a
simple change of the redshift when reionization occurs we can compare
$B2_{\rm cold}$ and $B2_{\rm hr}$ at $z=3$. Both simulations have
approximately the same values for $T_0$ and $\gamma$ at $z<4$ while
reionization occurs at $z\sim 6$ and $z\sim 17$, respectively. The
differences are of the order of 3\%, in agreement with the findings of
M04a.

\subsection{Finding a ``best guess'' model}
\label{bestguess}

In order to settle on a best guess model we have started with the coarse 
grid of hydro-simulations presented in VHS. We have complemented this
with simulations of a wider range of $\sigma_{8}$ and a wider range 
of thermal histories. 
We have fitted the flux power spectrum of all (60,400) simulations
listed in Table 1 to the SDSS flux power
spectrum allowing for errors in the correction to the data as
described in section \ref{datacorr}. We thereby leave the 
effective optical depth as a free parameter at all redshifts.

The last column in Table 1 gives the resulting $\chi_{\rm min}^2$
values obtained using the full covariance matrix as given 
by M04a {\footnote {\tt http://feynman.princeton.edu/$\sim$pmcdonal/LyaF/sdss.html}}. 
When performing the fits we noticed that the highest redshift bin at
$z=3.8$ is generally fitted very poorly. Omitting these 12 data points 
usually reduced the $\chi^2$  by about 26 for 11 degrees of freedom (d.o.f.). 
This may either indicate some problem with the data or
some insufficiency of the model. The latter would require, however, 
a very rapid change of the model with redshift as dropping any of the
other redshift bins reduced the  $\chi^2$ by the expected amount. 
UV fluctuations are probably the most plausible candidate for a rapid 
change toward the highest redshift bin.
We have decided to drop the highest redshift bin for our analysis.  
This leaves 96 data points and 8 free (unconstrained) parameters
corresponding to 88 degrees of freedom.  

The best fitting simulation appears to be B2.5 which has $\chi^2_{\rm
min} = 101.2$, when minimizing over the noise array $f_i$, the
resolution $\alpha$ and the effective
optical depth. The simulation B2 has a  $\chi^2_{\rm min}=104.7$
close  to that of B2.5, while the other models have typically $\Delta \chi^2
> 10$.  We regard these two models, with their thermal history, 
as reasonable good fits to the data set ($\chi^2$ values higher than the
quoted values   have a probability of 16\% and 11\% to occur).  

We decided to choose  B2 as our best guess model
despite the fact B2 has a somewhat larger $\chi^2$ 
value than B2.5 for two reasons. First,
the model is surprisingly close to the best fit model quoted by M04a
(which has $\Omega_{\rm m}=0.3$, $\sigma_8=0.85$, $n=0.94$ and $h=0.7$).
Secondly, we had already run more simulations exploring different thermal
histories for B2 than for the other models. 
We have, however,  checked as discussed in more detail in section
\ref{chifitting}  that an expansion around the B2.5 model gives essentially the same results
in terms of the inferred astrophysical and cosmological parameters.

\subsection{Derivatives of the flux power spectrum of the ``best guess'' model}
\label{derivatives}
\begin{figure*}
\center\resizebox{0.9\textwidth}{!}{\includegraphics{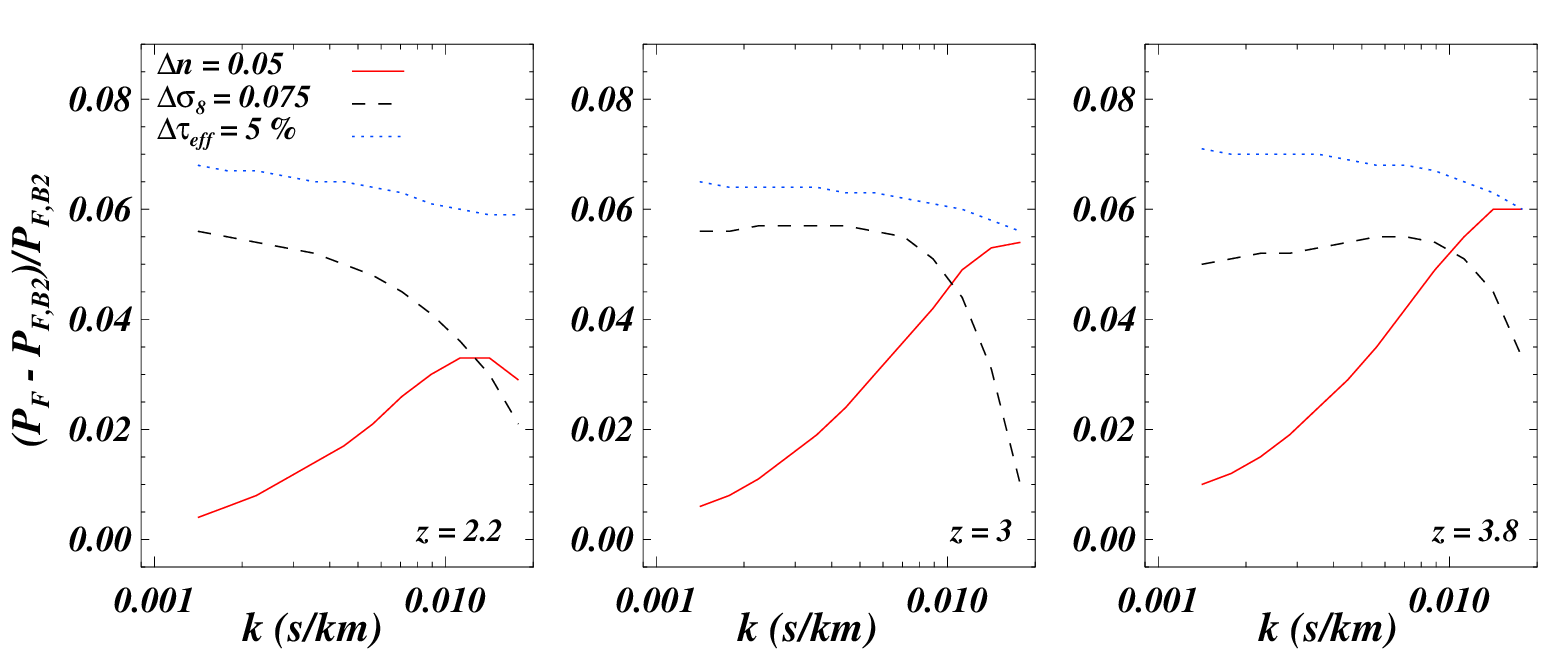}}
\center\resizebox{0.9\textwidth}{!}{\includegraphics{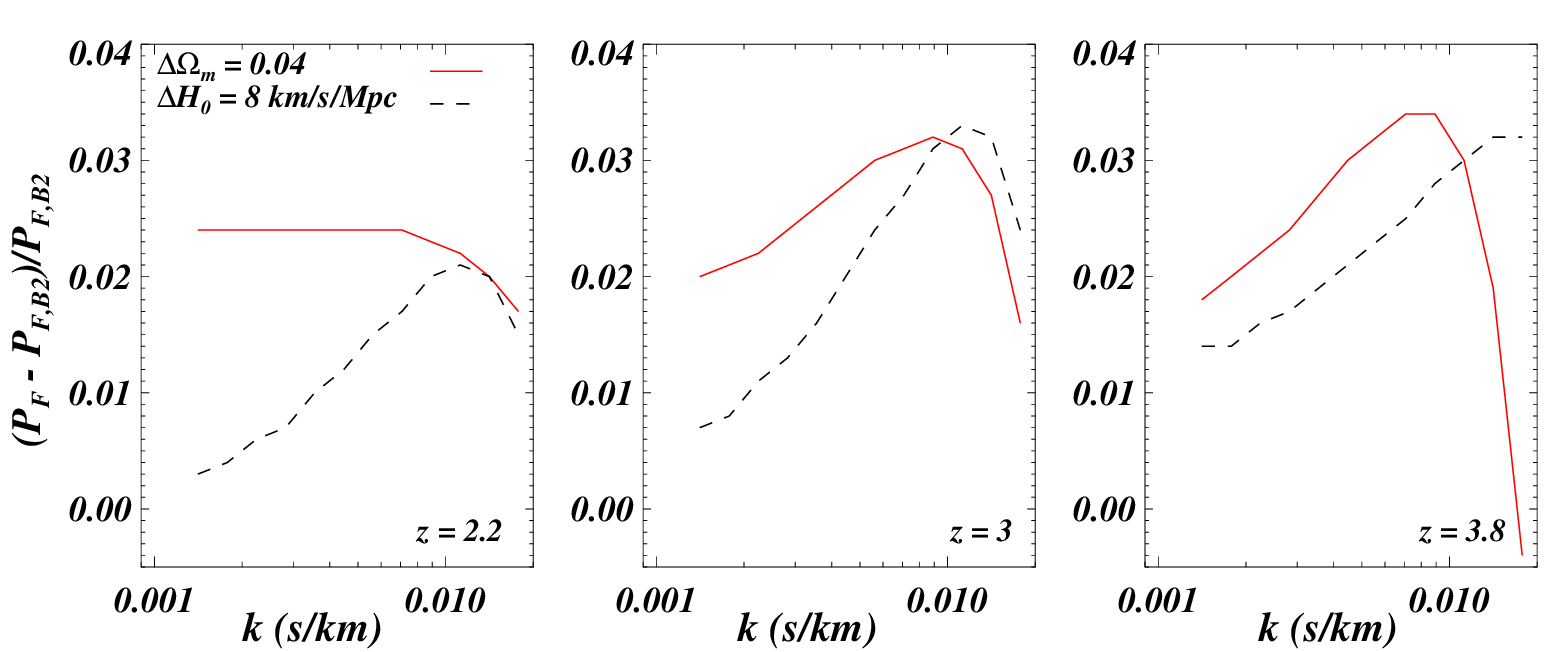}}
\center\resizebox{0.9\textwidth}{!}{\includegraphics{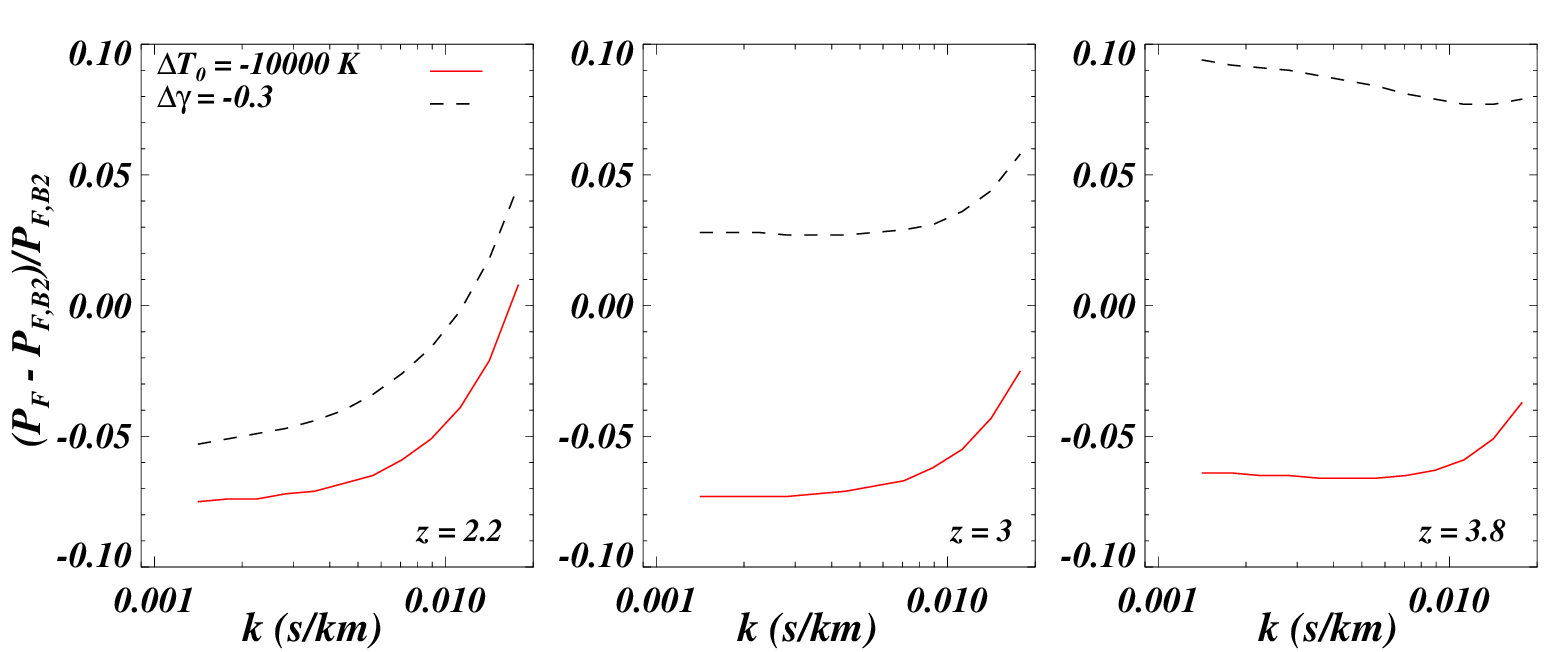}}
\caption{{\it Top}: Fractional differences in the flux power spectrum 
for variations of  $\sigma_8$ value (dashed curve), the spectral index $n$
(continuous curve) and the effective optical depth $\tau_{\rm eff}$
(dotted curve) at $z=2.2,3,3.8$ (left, center and right panel,
respectively). {\it Middle}: Fractional differences in the flux power
spectrum for variations of $\Omega_{\rm m}$ value (continuous curve), the
Hubble parameter $H_0$ (dashed curve). {\it Bottom}: Fractional differences in the flux
power spectrum for variations of $\gamma$  (continuous curve) and
$T_0$ (dashed curve).}
\label{f2}
\end{figure*}

The key ingredient for our estimates of a fine grid of flux power
spectra are the derivatives of the flux power spectrum 
(see equation \ref{taylor}) close to the
best guess model B2. Figure \ref{f2} shows
the $k-$dependence of the most relevant of these derivatives at
three redshifts ($z=2.2,3,3.8$ from left to right) in the form of the
change of the flux power spectrum for a finite change of the
parameters. We compute these derivatives at the wavenumbers 
of the SDSS flux power spectrum, although we cannot directly  
cover the two  smallest wavenumbers because we are using (30,200) 
simulations to estimate these derivatives. For these two points we 
have used an extrapolation with a 2nd order polynomial. This should be a reasonable
approximation given the rather weak and smooth $k-$dependence at these scales. 

In the top panel we show how the flux power
spectrum depends on  $\sigma_8$, $n$, $\tau_{\rm eff}$.  A $5\%$
increase in the effective optical depth, translates into a 7\%
increase in the flux power spectrum with no dependence on
redshift and with very little ($<1\%$) dependence on wavenumber.
A change of the spectral index by $\Delta n=0.05$
leads to an increase of the flux power which rises steeply 
with wavenumber from $1$ to $4-6\%$, while increasing $\sigma_8$ by 9\% 
increases the flux  power  by $5-6\%$ for $k <0.01$ (s/km) dropping to $\sim 2\%$ at
the smallest scales considered here.  The changes introduced by
different values of $\Omega_{\rm m}$ and $H_0$ are
shown in the middle panel. An increase of the Hubble
parameter by 8 km/s/Mpc leads to an increase of  order 3\%, while 
the change due to an increase of $\Omega_{\rm m}$ shows a peaks at $k\sim 0.01$
s/km of  3\%  at $z=3.8$ and depends less strongly on wavenumber at lower redshifts.
The derivatives of the flux power spectrum with
respect to changes in temperature and power-law index of the
temperature density relations are shown in the bottom panel. 
The trends are the same as those shown in Figure \ref{f3}. We note 
that the overall effect depends  
only weakly on wavenumber for $k<0.01$ s/km and steepens at smaller scales.
Note that the dependence of the derivatives on $k$ and $z$ is distinctively
different for the different parameters.  This is the reason why we 
will be able to obtain reasonably tight constraints for most of them.

Our results agree quite well with the results of VHS and M04b. We note,
however, that a direct quantitative comparison with M04b is not
obvious. Our best guess model has still somewhat different parameters
from their fiducial model and M04b have used HPM simulations with a
range of corrections while we have used full hydro-simulations.  One
may ask how much these derivatives depend on our choice of the
best-guess model.  We have checked this for a few cases.  The
derivatives obtained by rescaling the effective optical depth around
other models (B1,B3,C3) are different by less then 0.5 \% compared to
those shown in Figure \ref{f2}. The derivatives with respect to
$\sigma_8$ and $n$ differ by less than 2.5\% and 1.5 \%, respectively,
compared to expanding around B2.   We have also 
directly compared the approximate flux power obtained with the 
Taylor expansion with that extracted from hydrodynamical simulations
for a few simulation with parameters that are about$\magcir 2\sigma$ away 
from our best-guess model. For  the $\sigma_8=0.7$ and $\sigma_8=1$
(60,400) the error of the aproximation is less than  3.5\% which
should be compared to the difference to the best-guess model which 
is $\sim 10$\%.  We have also run a few further  (30,200) simulations 
with $n=1.05$, $\Omega_{\rm m}=0.34$ and a hotter simulations with 
$T_0$=41000 K and $\gamma=1.56$. The  approximations for the
models with different $\Omega_{\rm m}$ and the spectral index $n$ are accurate to 1\%. The 
error for the model with the different  $T_0$ parameter is $\sim 5$\%.
By comparing with Figure 3 we find that for $\magcir 2\sigma$ displacements in the
parameters $n$,$\sigma_8$,$\omega_{\rm m}$ and $T_0$ the error of 
the Taylor series approximation is generally less than  30\% of the 
difference between the models. This is not perfect but
should be acceptable considering the expense of full-hydrodynamical
simulations and the size of the parameter space.

\section{Constraining Astrophysical and Cosmological Parameters}
\label{chifitting}

\subsection{Summary of the free parameters of the $\chi^2$ minimization}    

We have modified the code COSMOMC (Lewis \& Bridle 2002) to run Monte Carlo
Markov Chains for our set of parameters.  After settling on a
best guess model we have explored the parameter space close to the best
guess model. For most of the analysis we have used 22 parameters for
the $\chi^2$ minimization, some of them free some of them independently
constrained.  We briefly summarize the parameters here.

We assume a flat cosmological model with a cosmological constant and
neglect the possibility of a non-zero neutrino mass. There are thus
four cosmological parameters that describe the matter distribution: the
spectral index $n$, the dark matter power spectrum amplitude
$\sigma_8$, $\Omega_{\rm m}$ and $H_0$.  We use nine ``astrophysical''
parameters. Two describe the effective evolution of the optical depth
evolution, six describe the evolution of $\gamma$ and $T_0$ 
and one describes the contribution of
strong absorption systems. For the evolution of the optical depth we
assume a power-law $\tau_{\rm eff}(z) = \tau_{\rm
eff}^A(z=3)[(1+z)/4]^{\tau_{\rm eff}^S}$. Note that this is different
from what we did when we fitted the suite of hydro-simulations where we
let the effective optical depth vary independently at all redshift bins. We
will come back to this in section \ref{optical}.

For the
evolution of $\gamma$ and $T_0$ we assume broken power-laws with a
break at $z=3$.  The three parameters for the temperature are the
amplitude at $z=3$, $T_0^A (z=3)$ and the two slopes $T_0^S(z<3)$ and
$T_0^S(z>3)$. $\gamma$ is described in the same way by $\gamma^A
(z=3)$, $\gamma^S(z<3)$ and $\gamma^S(z>3)$. The correction for the
damped/high column density systems is modelled with the $k-$dependence
of Figure 11 of M04b and an overall amplitude $A_{\rm damp}$ and no
redshift evolution, as described in section \ref{moderrors}.

Finally, we have a total of 9 parameters which 
model uncertainties  in the correction to the data. Eight parameters 
for the noise correction $f_{i}$ as described in section \ref{datacorr}
and one  parameter $\alpha$ describing the error of the
resolution correction.  Note that all the parameters describing the errors
of the corrections to the data and the correction for damped/high
column density  systems  are constrained as suggested in M04b and described in section
\ref{system}. 

We will present here two sets of results. At first we will discuss the 1D
and 2D likelihoods for the most significant parameters by assuming no
priors on the cosmological and astrophysical parameters
(with the exception of the correction for damped systems). 
Then we will show some results  assuming priors on the 
Hubble parameter and on the thermal history.

\begin{figure*}
\center\resizebox{1.0\textwidth}{!}{\includegraphics{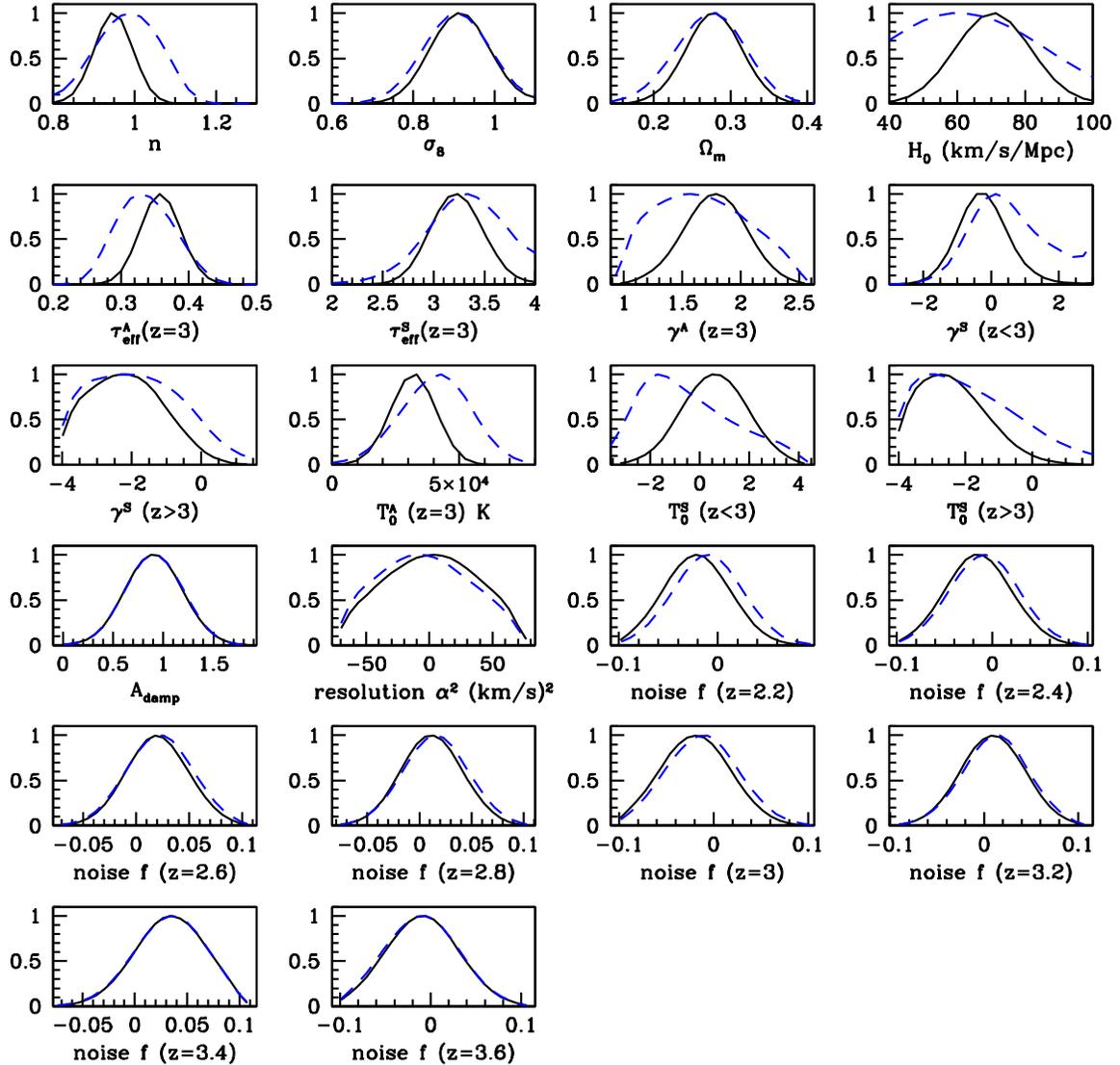}}
\caption{1D marginalized likelihoods for our 22 parameters. Dashed curves
represents the case without priors, while the continuous curves are
obtained with the priors discussed in section \ref{priors}.
Cosmological and astrophysical parameters were inferred with
a Taylor expansion of the flux power spectrum of the 
best-guess model to first order.}
\label{f4}
\end{figure*}

\begin{figure*}
\center\resizebox{1.0\textwidth}{!}{\includegraphics{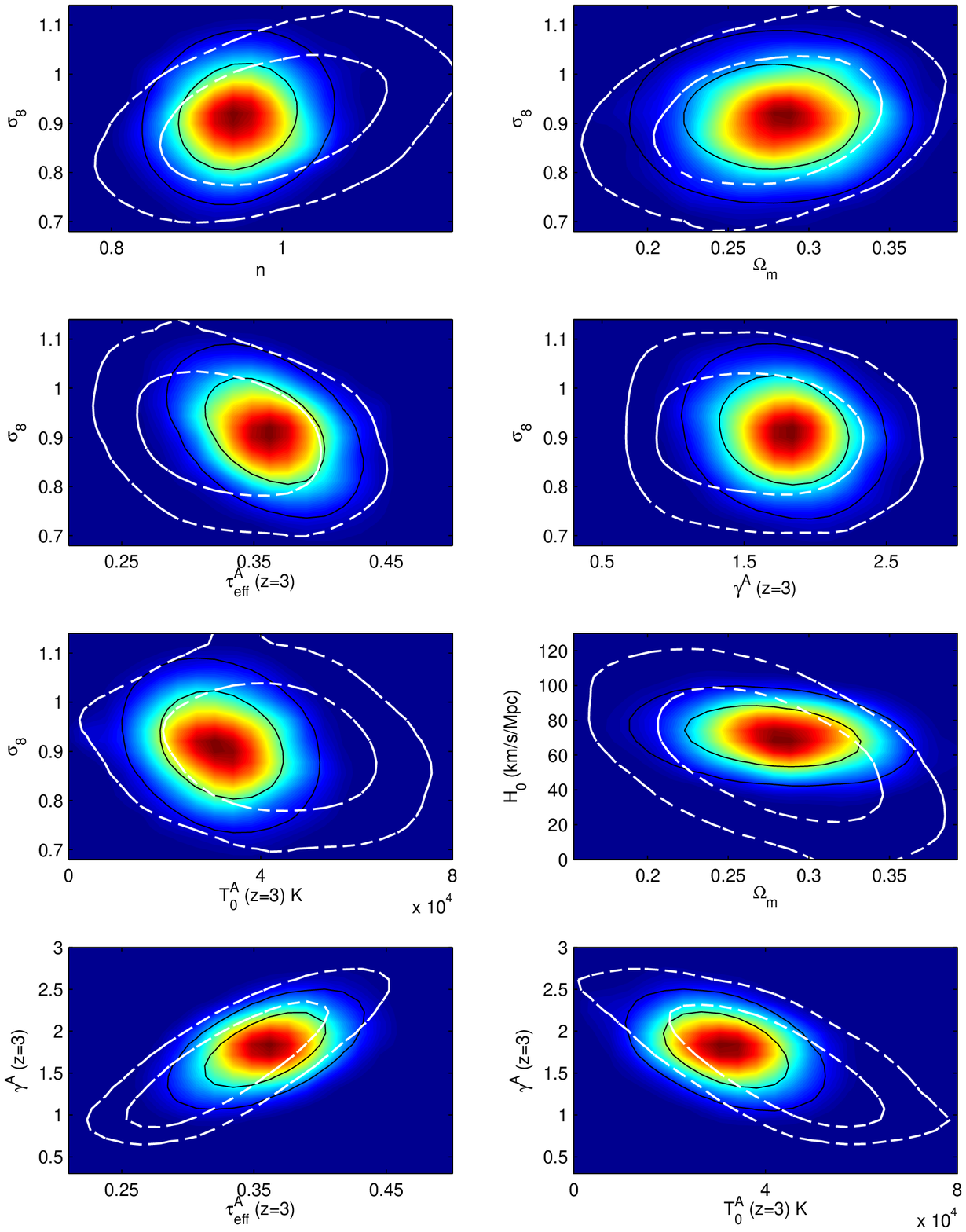}}
\caption{2D likelihoods for some of the parameters used in our
analysis. Filled (coloured), continuous and white dashed contours refer
to the mean likelihood for the case with priors, the marginalized
likelihood for the case with priors and the marginalized likelihood for
the case without priors, respectively. Cosmological and astrophysical
parameters were inferred with a Taylor expansion of the flux power
spectrum of the best-guess model to first order.}
\label{f5}
\end{figure*}

\subsection{Results without priors} 
\label{nopriors}

The dashed curves in Fig. \ref{f4} show the 1D (marginalized) likelihoods
for our full set of parameters without imposing any priors on the
cosmological and astrophysical parameters.   The best fit model has 
now  $\chi^2_{\rm min}=78.9$ for 84  d.o.f.. A $\chi^2$  
equal or larger than this should have a probability of 
64\%. Our best-fitting model is thus an excellent fit, maybe 
with a hint that we are already marginally over-fitting 
the  data.  Most parameters including
$\sigma_8$, $n$ and $\Omega_{\rm m}$ are tightly constrained.  The
constraints on the thermal history are considerable weaker.  This is
not surprising as the rather low resolution of the SDSS spectra means
that the thermal cut-off at small scales is not actually resolved.  The
best fit value for $T_0^A$ is rather high, $4.1\pm 1.3 \times 10^4$ K,
 and may be in  disagreement with the best fit values of M04b who find 
$T_0^A=2 \times 10^4$ K without giving any error estimate.

The dashed curves in Figure \ref{f5} show 2D likelihood contours for
various parameter combinations. The contours have been obtained by
marginalizing over the rest of the parameters.  We note that the best
fit models prefer somewhat larger values of the spectral index and
higher temperatures than those of our best guess model. The 2D contour
plot of $\gamma^A$ vs $T_0^A$, suggests that in order to reconcile this
measurement with a colder IGM one would need an unplausibly steep
temperature density relation with $\gamma \sim 2$.

The preferred values for the thermal state of the IGM are actually outside the range
suggested by a detailed analysis of the line width of the absorption
feature using high-resolution spectra.  This is shown in Figure \ref{f6}
where we compare the best-fitting evolution of the temperature $T_0$
(hashed region) to the measurements of the temperature by Schaye et
al. (2000). Note again that the lower range of values corresponds to
$\gamma\magcir 2$  outside of the range of plausible values for a
photoionized IGM. The Hubble constant is  also only poorly
constrained.  This is again not surprising as
the spectra are sensitive to the matter power spectrum in velocity
space.  The only dependence comes thus through the shape parameter
$\Gamma = \Omega_{\rm m} h$, which is weak (\ref{f2}).  

The best fitting values for our modelling
without priors with errors obtained by marginalizing over all other
parameters are listed in  Table 2.

As a further consistency check and to test how well the 
observed redshift evolution of the flux power spectrum probes the
expected gravitational growth of structure we have also obtained  
constraints on  $\sigma_8$ and $n$ independently for  each redshift
bin.  We get the
following constraints for $\sigma_8$: $0.92\pm 0.16$, $0.89 \pm 0.11$,
$0.98 \pm 0.13$, $1.02 \pm 0.09$, $0.94 \pm 0.08$, $0.88 \pm 0.09$,
$0.87\pm 0.12$, $0.95 \pm 0.16$, $0.90 \pm 0.17$, for
$z=2.2,2.4,2.6,2.8,3,3.2,3.4,3.6,3.8$, respectively. For $n$ we get:
$0.88 \pm 0.09$, $0.97 \pm 0.07$, $0.89 \pm 0.08$, $0.90 \pm 0.06$, $0.90 \pm
0.06$, $0.94 \pm 0.07$, $0.99 \pm 0.06$, $0.90 \pm 0.05$, $0.95\pm
0.14$ for $z=2.2,2.4,2.6,2.8,3,3.2,3.4,3.6,3.8$, respectively.
The total $\chi^2_{\rm min}$ is $77.9$.  for 74 d.o.f. in this case. 
The constraints are weaker but perfectly consistent with our 
estimates from all redshift bins combined. The errors are significantly
smaller than the expected growth of the amplitude between  the lower
and upper redshift end of the sample demonstrating  that the flux
power spectrum evolves as expected for gravitational growth.


\begin{table}
\caption{Cosmological and astrophysical parameters inferred with
a Taylor expansion of the flux power spectrum of the 
 best-guess model to first order(1$\sigma$ error bars)}
\label{tab2}
\begin{tabular}{lcc}
parameter & without priors &  with priors on thermal \\
                          &&  history and $H_0$\\ 
\hline
\noalign{\smallskip}
  n               & $0.98 \pm 0.07 $  & $0.95 \pm 0.04 $ \\
  $\sigma_8$         & $0.90 \pm 0.08 $ & $0.91 \pm 0.07$  \\
  $\Omega_{\rm m}$        & $0.27 \pm 0.04$  &  $0.28 \pm 0.03 $\\
  $H_0$ (km/s/Mpc)	  & $62\pm20$  & $70\pm10$ \\
  $\tau^A_{\rm eff} (z=3)$ & $0.337 \pm 0.040 $ & $0.359 \pm 0.027 $ \\
  $\tau^S_{\rm eff}$ & $3.29 \pm 0.36 $ & $3.22 \pm 0.24 $ \\
  $\gamma^A (z=3)$         &  $1.67 \pm 0.38 $ & $1.78 \pm 0.26 $\\
  $\gamma^S (z<3)$   &  $0.71 \pm 1.26 $ & $-0.21 \pm 0.74 $  \\
  $\gamma^S (z>3)$   &  $-1.83 \pm 1.03 $ &  $-2.15 \pm 1.00 $\\
  $T_0  (z=3) (10^4)$ K   &  $4.1 \pm 1.3 $ &  $3.2 \pm 0.8 $\\
  $T_0^S (z<3)$      &  $-0.49 \pm 1.83 $ &$0.60 \pm 1.27 $\\
  $T_0^S (z>3)$      &  $-1.78 \pm 1.58 $ &$-2.38 \pm 1.01 $\\
  $A_{\rm damp}$             &  $0.91 \pm 0.28  $ &$0.91 \pm 0.27  $\\
\hline
\noalign{\smallskip}
\end{tabular}

\end{table}

\subsection{Results with priors}
\label{priors}

As discussed in the last section the SDSS flux power spectrum of 
SDSS published by M04a prefers unplausibly  large temperatures 
for reasonable values of $\gamma$.  We have thus repeated the 
analysis with  (Gaussian) priors on the thermal state of the IGM, 
as follows, $T_0^A(z=3)=2.1\pm0.7\times 10^4$,
$T_0^S(z<3)=1.45\pm1.00$ and $T_0^S(z>3)=-2.44\pm 1.00$ (Schaye et
al. 2000). We have also added a prior on the Hubble constant, 
$H_0=72\pm 8$ km/s/Mpc (Freedman et al. 2001).

\begin{figure*}
\center\resizebox{0.5\textwidth}{!}{\includegraphics{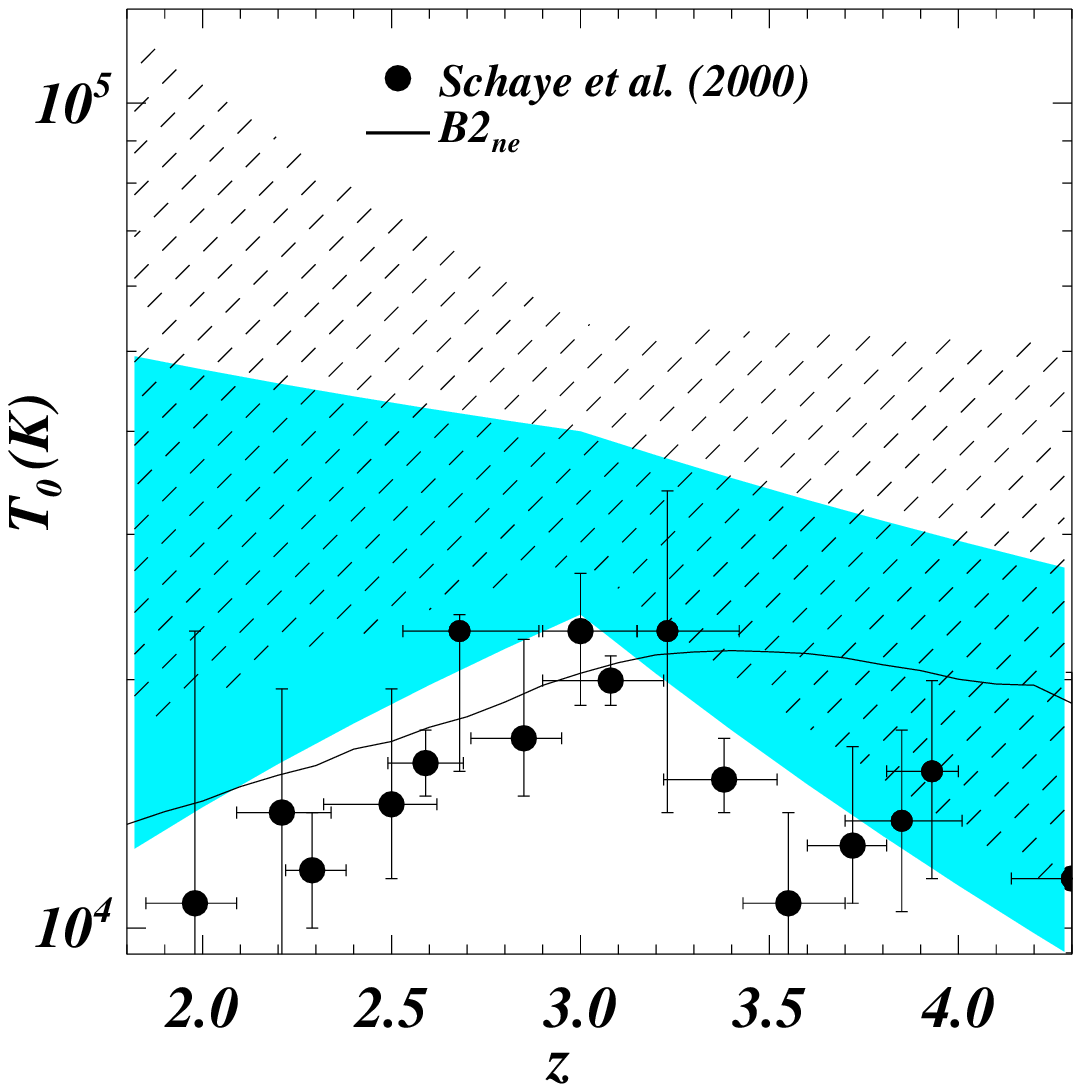}}
\caption{Evolution of the temperature at the mean density $T_0$ with
redshift inferred for our analysis without priors (hashed area) and 
with priors (shaded area). The regions indicate $1\sigma$
uncertainties. The observational points by Schaye et al. (2000)
and the thermal evolution of our simulation of model B2 
with the non-equilibrium solver are also shown.}
\label{f6}
\end{figure*}

The solid curves 
in Figure \ref{f4} and Figure \ref{f5} show the 
marginalized likelihoods for our analysis with priors. 
The  filled (coloured) contours 
in Figure \ref{f5} show the  mean  2D likelihood.
As expected the thermal parameters 
and the Hubble constant are now more tightly constrained. 
The temperature evolution of the best-fitting model with 
prior is shown as the shaded  region in Figure \ref{f6}. 
It lies about 1$\sigma$  above the imposed constraint suggesting
that the observed flux power spectrum definitely prefers 
models with higher temperatures than those observed.  Rather than being an actual
indication of higher temperature this is more likely indicative 
of a not understood systematic uncertainty in our
modelling which makes the flux power spectrum mimic high
temperatures. As discussed in section \ref{othereffects} there are plenty of
candidates for this.

The best fit model with priors has  $\chi^2_{\rm min}=83.1$ 
For 88 degrees of freedom.  A $\chi^2$  
equal or larger than this should have a probability of 
60\% very similar to the case without priors. 
The best fitting values for our modelling without priors are 
also listed in  Table 2 (right column). 

The most significant change caused by the introduction of the priors is
a decrease of the spectral index by $\Delta n=0.03$ (consistent with
the $1\sigma$ errors). In Figure \ref{f5} the
strongest correlations are those in the $\sigma_8-\tau_{\rm eff}^A$, $\gamma^A-T_0^A$ and
$\gamma^A-\tau_{\rm eff}$ planes.  As expected a higher value of 
the $\tau_{\rm eff}^A$ requires a smaller $\sigma_8$  quantitatively 
comparable to previous findings  (Viel, Weller \& Haehnelt 2004; VHS; Seljak et al. 2003).
The best-fitting  temperature and slope of temperature density
relation are again anti-correlated. 
A lower temperature corresponds to a steeper
temperature density relation.  We further note that $\sigma_8$ and $n$
appear not to be correlated, in contrast to the corresponding 
parameters $(\Delta^2_L,n_{\rm eff})$ in M04b.
If we take out the box size and resolution corrections we do
not get significant changes in the final parameters but the
$\chi^2_{\rm min}$ increases by 1.3, showing that the data prefer these
corrections.
The best fitting values for our modelling with priors with errors 
obtained by marginalizing over all other parameters are also listed in
Table 2.

\subsection{The inferred evolution of the effective optical depth}
\label{optical}
\begin{figure*}
\center\resizebox{1.0\textwidth}{!}{\includegraphics{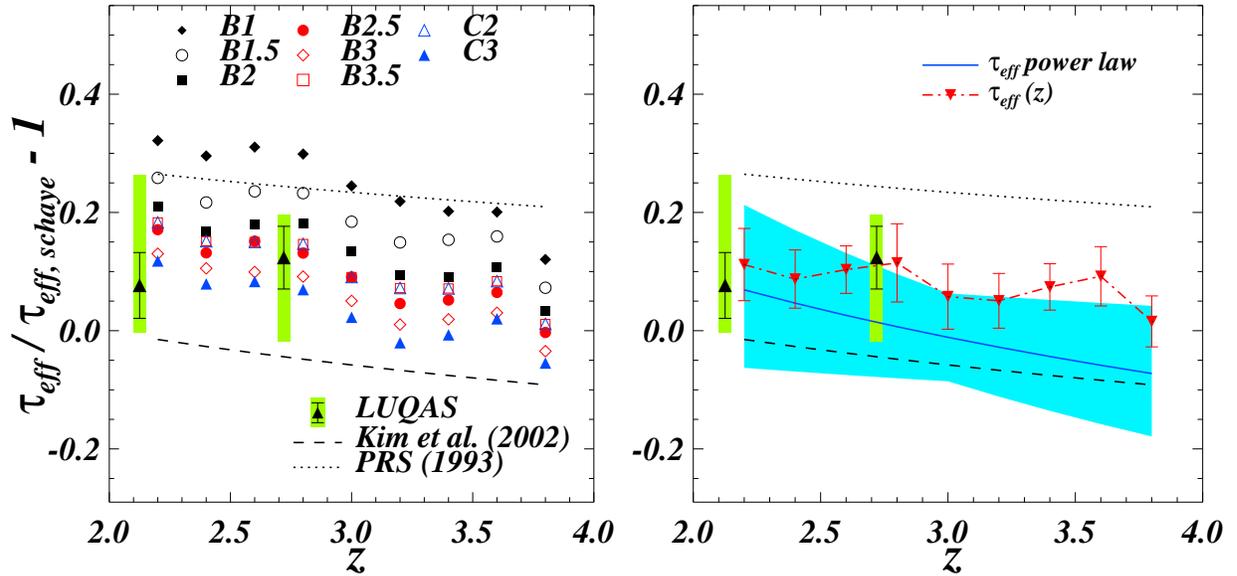}}
\caption{{\it Left}: fractional differences of the best fitting value
of the effective optical depth with respect to the evolution found by
Schaye et al. (2003) for some of the models of Table 1 (with the
effective optical depth in the eight redshift bins as the only free
parameters).  The observed values obtained for the LUQAS sample are
represented by triangles, while the shaded bars indicate the range used
in the analysis by VHS. The dotted line and the dashed line show
the observed evolution found by Kim et al. (2002) and Press, Rybicki \&
Schneider (1993), respectively. {\it Right}: same as in the left panel but
using the flux power spectra calculated using a Taylor expansion to
first order around the ``best guess'' model B2. For the the continuous
curve the effective optical depth was assumed to evolve as a power-law
with redshift while the triangles show the case of leaving the optical
depth float freely in all redshift bins.}
\label{f7}
\end{figure*}

As discussed before the effective optical depth is the largest
uncertainty and it is very degenerate with the fluctuation
amplitude. Larger effective optical depths correspond to smaller
$\sigma_{8}$ values (Figure \ref{f5}). The shape of the derivative of
the flux power spectrum with respect to $\tau_{\rm eff}$ and
$\sigma_{8}$ (Figure \ref{f2}) is, however, sufficiently different to get
still interesting constraints on $\tau_{\rm eff}$.

In the left panel of Figure \ref{f7} we compare the best fitting $\tau_{\rm
eff}$ for our set of (60,400) hydrodynamical simulations to
measurements made from high and low resolution spectra.  
All values are shown relative to the fit to the observed evolution of 
the optical depth corrected for the contribution by associated 
metal absorption based on a set of high-resolution spectra 
by Schaye et al. 2003 (their Figure 1).  The fit is given by 
$\tau_{\rm eff}=0.363\,([(1+z)/4]^{3.57}$.  The evolution obtained by
Kim et al. (2002) based on a different set of high-resolution spectra  
is shown as the dashed curve. The measurements at $z=2.125$ and
$z=2.72$ made by VHS from the LUQAS sample (Kim et al. 2004) are 
shown as filled triangles.   Note that  there is some overlap between the 
Schaye et al. (2003) data set and the 
LUQAS sample of K04.   The dotted line shows the result of  Press, 
Rybicki \& Schneider (PRS, 1993) from low-resolution spectra
of rather poor quality.  The shaded area shows the range adopted by VHS and Viel, 
Weller \& Haehnelt (2004) for their analysis.  
The different symbols show the results from our fits
of the flux power spectrum of the hydro-simulations to the SDSS flux
power spectrum as described in section \ref{bestguess}. The degeneracy 
between $\sigma_8$ and $\tau_{\rm eff}$ is clearly visible
and  can be  roughly described by $\Delta \tau_{\rm eff}/\tau_{\rm eff}=0.8\,\Delta\sigma_8$  
similar to that  found in VHS.   

In the right panel, the solid curve shows the evolution of $\tau_{\rm eff}$
for our best fitting model compared to the same observational
estimates.  As discussed we have modelled the evolution of $\tau_{\rm
eff}$ as a power-law. We find $\tau_{\rm eff}(z) = (0.359\pm0.027)
\,[(1+z)/4]^{3.22\pm0.24}$ in agreement with Schaye et
al. (2003) and the values used by VHS. It also agrees well with the
values inferred  by Lidz et al. (2005) from the flux probability
distribution.  The shaded area in the right
panel shows the $1\sigma$ uncertainty of the best fitting model.
We thus confirm the findings of M04b that the differences in the
$k-$dependence of the respective changes can break the degeneracy between
$\sigma_8$ and $\tau_{\rm eff}$. 
The dot-dashed line and the triangles show the evolution of the
optical depth if we let $\tau_{\rm eff}$ vary separately at all nine  
redshifts (this increases the number of free parameters from 22 to 28).  
Letting the 
effective optical depth vary freely at high redshift 
increases the average inferred effective optical depth by about 5\% 
and reduces the  inferred errors.

\begin{table*}
\caption{Cosmological parameters obtained in recent studies of the \lya forest}
\label{tab3}
\begin{tabular}{lcccc}
\hline
\noalign{\smallskip}
& $\sigma_8$ & $n$ \\
\noalign{\smallskip}
\hline
\noalign{\smallskip}
Spergel et al. (2003)$^{\mathrm{(a)}}$ & $0.9 \pm 0.1$   & $0.99 \pm 0.04$ \\
Spergel et al. (2003)$^{\mathrm{(b)}}$ & --   & $0.96 \pm 0.02$ \\
Viel, Haehnelt \& Springel (2004)$^{\mathrm{(c)}}$  & $0.93 \pm 0.10$   & $1.01 \pm 0.06$\\
Viel, Weller \& Haehnelt (2004)$^{\mathrm{(d)}}$   & $0.94 \pm 0.08$ & $0.99 \pm 0.03$ \\
Desjacques \& Nusser (2004)    & $0.90 \pm 0.05$ & -- \\
Tytler et al. (2004)       & 0.90 & -- \\
McDonald et al. (2004b)$^{\mathrm{(e)}}$ 
& $0.85 \pm 0.06$ & $0.94 \pm 0.05$ \\
Seljak et al. (2004)$^{\mathrm{(f)}}$ & $0.89 \pm 0.03$ & $0.99 \pm 0.03$ \\
This work (priors case)     & $0.91 \pm 0.07$ & $0.95 \pm 0.04$ \\
\hline
\noalign{\smallskip}
\end{tabular}
\begin{list}{}{}
\small
\item[$^{\mathrm{(a)}}$ WMAP only, $^{\mathrm{(b)}}$ WMAPext +2dF +
\lya (no running of the spectral index),$^{\mathrm{(c)}}$] LUQAS+Croft et
al. (2002) only; $^{\mathrm{(d)}}$ LUQAS+Croft et al. (2002) + WMAP 1st
year; $^{\mathrm{(e)}}$ error bars extrapolated by the authors from the
abstract of M04b; $^{\mathrm{(f)}}$ SDSS \lya flux power spectrum + WMAP.
\end{list}
\end{table*}

\section{Discussion and Conclusion}
\label{conclu}

\subsection{Comparison with previous results for $\sigma_8$ and $n$  }

In Table \ref{tab3} we list results for the most important parameters
describing the matter distribution ($\sigma_8$ and $n$) obtained by
recent studies using the \lya forest by a variety of authors. The most
direct comparison is again with M04b. At $z=3$ 
our results correspond to an fluctuation amplitude and effective slope 
at the pivot wavenumber $k=0.009$ s/km as defined by M04b of 
$\Delta^2_L=0.40 \pm 0.06$ and $n_{\rm eff}=-2.33\pm0.05$.  The 
values are consistent within $1\,\sigma$ with those of M04b.  
The agreement with otherauthors including our own work is at the same level. For comparison we
also show the value for WMAP alone.  A consistent picture emerges for
the \lya forest data with a rather high fluctuation amplitude and no
evidence for a significant deviation from $n=1$ or a running of the
spectral index.  This results holds for the \lya forest data alone but
is considerably strengthened if the \lya forest data is combined with
the CMB and other data.

\subsection{Remaining Uncertainties and Future Progress}

As discussed in section \ref{system} there are many systematic
uncertainties that affect the measurement of cosmological parameters
with the \lya flux power spectrum.

In order to test the stability of the results we have removed the
correction  for high column density systems 
and the errors for the noise and resolution corrections in
the case with priors.  
If we take out the correction due to strong absorption systems $n$,
decreases  to $0.93 \pm 0.04$, and the best fit value for $\tau_{\rm
eff} (z=3)$ becomes 0.381, all the other parameters remain practically
unchanged. The measured values do therefore practically not
change but the value of $\chi^2_{\rm min}$ increases by 0.5.
If we do not allow for the error to the noise  and  resolution
corrections the inferred values do again not change 
but the value of  $\chi^2_{\rm min}$ increases by 1.8.
Note that M04a have also subtracted an estimate of the contribution 
of associated metal absorption from the flux power spectrum and have 
applied a suite of other (smaller) corrections to the data 
(see M04a for more details). Many of these
corrections would either not be necessary or substantially easier for
spectra with somewhat better resolution ($R\sim 10000$) and higher
S/N. Obtaining large samples of such spectra is observationally
feasible and there is certainly room for improvement.
Note further that the SDSS team will soon release a further 
measurement of the flux power spectrum performed by an independent
group within the collaboration. 

We also note again that calculating the flux power spectrum by using a
Taylor expansion to first order around the best guess model is just an
approximation. This was an important part of our analysis as it reduced
the amount of CPU time necessary dramatically, but could be further
improved by getting more accurate fits around the ``best guess'' model.
We have also not attempted to correct for the possible effects of
spatial fluctuations of the \H ionization rate, galactic winds,
temperature fluctuation or the reionization history. This is because we
are not convinced that it is already possible to model these effects
with sufficient accuracy. This should hopefully change in future with
improved observational constraints and numerical capabilities.  One
should further keep in mind that so far very little cross checking of
hydrodynamical simulation run with different codes has been
performed.

In our study the rather poorly known thermal state
of the IGM is one of the major remaining uncertainties.  Further
high-resolution spectroscopy and improved modelling will hopefully
soon improve this. Accurate modelling of the effect of
high column density systems/ damped \lya systems and improvement in the
determination of the column density distribution of observed absorption
systems in the poorly determined range around $\log N \sim 17$ should
also be a priority in further studies (see the discussion in McDonald
et al. 2005).

It is certainly encouraging that the differences between our analysis
and that of M04b are moderate despite considerable differences between
their approximate HPM simulations calibrated with hydrodynamical
simulations of rather small box size and our analysis with 
full hydrodynamical simulations of much larger box size here and in 
VHS.

For the time being we would advice the conservative reader  to double 
the formal errors quoted here. This will bring the error estimates 
to  about the same size as the also conservative estimates 
for the errors of the fluctuation amplitude from the \lya forest 
data alone in the analysis of  VHS.  The actual errors 
lie probably somewhere in between.

\subsection{Conclusions}

We have compared the flux power spectra calculated from a suite of full
large box-size high resolution ($60h^{-1}$ Mpc, $2\times 400^3$
particles) hydrodynamical simulations with the SDSS
flux power spectrum as published by McDonald et al.  (2005).  We have
identified a best-guess model which provides a good fit to the data.
We have used a Taylor expansion to first order to calculate flux power
spectra in a multi-dimensional space of parameters describing the
matter power spectrum and the thermal history of the IGM with values
close to those of our best-guess model.  We have 
investigated the combined effect of cosmological and astrophysical
parameters on the flux power spectrum with an adapted version of the
Markov-Chain code COSMOMC.  Our main results can be summarized as
follows.
\begin{itemize}
\item{The flux power spectrum calculated directly from the simulation
of a $\Lambda$CDM model ($\Omega_{{\rm m}}= 0.26$, $\Omega_{\Lambda} =
0.74$, $\Omega_{{\rm b}} = 0.0463$ and $H_0=72\,{\rm
km\,s^{-1}Mpc^{-1}}$,$\sigma_8 = 0.925$ ) with a temperature density
relation described by $T_0(z=3)=21000$ and $\gamma(z=3)=1.6$ gives an
acceptable fit to the SDSS flux power spectrum in the redshift range
$2.2<z<3.6$ ($\chi^2 \sim 101 $ for 88 degrees of freedom).  The fit can
easily be further improved by small changes in the cosmological and
astrophysical parameters. }
\item{At higher redshift the deviations from the observed flux power
spectrum become significantly larger ($\Delta \chi^{2} = 26$ for 12
additional data points at $z=3.8$) suggesting either some problem with
the data or a physical effect that changes rapidly with redshift.}
\item{We confirm the claim by McDonald et al. (2004) that the
degeneracy of the dependence of the flux power spectrum on the
amplitude of the matter power spectrum and the effective optical depth
can be broken for the published SDSS flux power spectrum. It will be
interesting to see if the same is true for the independent analysis of
the SDSS data to be released soon.}
\item{The SDSS power spectrum alone can constrain the amplitude of the
matter power spectrum, the matter density and the power-law index of
primordial density fluctuation to within $\sim5-10\%
$. The thermal state of
the IGM is however, poorly constrained and the SDSS power spectrum
formally prefers models with unplausible values of the parameters
describing the thermal state. The dependence of the flux power spectrum
on the assumed Hubble constant is also very weak and the Hubble
constant is not well constrained.  The exact values for the other
cosmological parameters depend somewhat on the assumed prior for the
thermal state and the details of the correction which have been applied
to the data. }
\item{With a prior on the thermal history and the Hubble constant
motivated by the observations by Schaye et al. (2000) and Freedman et
al. (2001) we obtain the following best-fitting values for the
cosmological parameters  $\Omega_{\rm m}=0.28 \pm 0.03$,
$n=0.95\pm0.04$, $\sigma_8=0.91\pm0.07$ ($1\sigma$ error bars), and the
effective optical depth is well described by the following power-law
relation: $\tau_{\rm eff}(z) = (0.359\pm0.027)
\,[(1+z)/4]^{3.22\pm0.24}$.  The
errors were obtained by marginalizing over a set of 22 parameters
describing the matter distribution, thermal history of the Universe,
the effective optical depth and errors to various corrections to the
data. The values for $\sigma_{8}$ and $n$ are consistent with those
found by M04a for the same data set with different
simulations. They are also consistent with the results of other recent
studies of \lya forest data. The inferred optical depth is in good
agreement with that measured directly from continuum-fitted
high-quality absorption spectra.}
\end{itemize}

\section*{Acknowledgments.} 
The simulations were run on the COSMOS (SGI Altix 3700) supercomputer
at the Department of Applied Mathematics and Theoretical Physics in
Cambridge. COSMOS is a UK-CCC facility which is supported by HEFCE and
PPARC. MV thanks PPARC for financial support. We thank James Bolton for
providing us with the non-equilibrium solver for \gad, Antony Lewis
for useful suggestions and technical help and the anonymous referee for
useful suggestions.

\end{document}